# Terahertz Electrodynamics and Kinetic Inductance of Disordered Titanium-Vanadium Alloy Thin Films

Shekhar Chandra Pandey[1,2*], Shilpam Sharma[1], Ashish Khandelwal[1] and M. K. Chattopadhyay[1,2]

[1] *Free Electron Laser Utilization Laboratory, Raja Ramanna Center for Advanced Technology, Indore, Madhya Pradesh - 452013, India*

[2] *Homi Bhabha National Institute, Training School Complex, Anushakti Nagar, Mumbai 400 094, India*

*Email- shekharpandey7579@gmail.com

**Abstract**. Disordered superconductors represent an important area in modern condensed matter physics, where superconductivity survives even in the presence of strong electron scattering and localization effects. Understanding how disorder modifies the high-frequency electrodynamic response is not only important from physics point of view, but is also essential for developing next-generation quantum detectors and superconducting devices. In this work, we investigate the terahertz electrodynamics of disordered $Ti_{40}V_{60}$ alloy thin films using terahertz time-domain spectroscopy (THz-TDS) to understand the relationship between disorder, quasiparticle dynamics, and kinetic inductance. By analysing the complex conductivity, penetration depth and superfluid response, we show that structural disorder can be systematically used to tune the inductive response while maintaining a robust superconducting phase. Unlike conventional nitride superconductors that require tightly controlled reactive growth conditions, $Ti_{40}V_{60}$ alloys provide a simpler and more adaptable route for tuning the superconducting energy scales directly through the deposition conditions. These findings establish $Ti_{40}V_{60}$ alloys as a promising material for kinetic inductance detectors and provide useful insights into the electrodynamics of strongly disordered superconductors.

Disordered superconducting thin films are foundational to the next-generation quantum sensors and detectors operating across the microwave to infrared spectrum [1-3]. Superconducting detectors such as kinetic inductance detectors (KIDs) and superconducting nanowire single-photon detectors (SNSPDs), rely on the superconducting electrodynamic response [4,5]. In highly disordered systems, enhanced electron scattering modifies the complex conductivity and increases the kinetic inductance ($L_K$), which is important for achieving high responsivity in the resonator-based devices [4]. Conventional nitride superconductors [1-3,6-8] and amorphous silicides [9-12] are widely used for such applications because of their high normal-superconducting transition temperatures ($T_C$s) and large $L_K$. Among the emerging alternatives, the Ti-V alloy thin films are attractive because they combine structural disorder, robust superconductivity, and high normal-state resistivity [13-16]. In addition, their superconducting properties can be tuned directly through the deposition conditions, avoiding the strict growth control required for the reactive deposition technique of the nitride superconductors [13-15].

The influence of disorder on the high-frequency electrodynamics of Ti-V alloys remains largely unexplored. THz-TDS provides a direct and non-contact method to probe the complex conductivity and superconducting quasiparticle dynamics [4,17,18]. In this Letter, we investigate the THz electrodynamics of $Ti_{40}V_{60}$ alloy thin films deposited under two different deposition pressures. By analyzing the temperature-dependent complex conductivity, important superconducting parameters such as the penetration depth ($\lambda$), superfluid density ($n_s$), superfluid stiffness ($J$), and kinetic inductance ($L_K$) have been extracted. The measurements reveal a strong superconducting inductive response together with clear signatures of disorder-driven electrodynamics in the $Ti_{40}V_{60}$ films. Our results show that deposition-induced disorder can be used to tune the superconducting electrodynamics and $L_K$ of $Ti_{40}V_{60}$ films for quantum device applications.

Two samples of $Ti_{40}V_{60}$ alloy thin films (20 nm thick), denoted as TiV-1 and TiV-2, were deposited on 500 $\mu$m thick silicon substrates (coated with 300 nm $SiO_2$) via DC magnetron co-sputtering at 0.7 and 0.63 $\mu$bar deposition pressures respectively, so that they have slightly different disorder levels. Comprehensive structural and DC electrical characterization of similarly grown films have been reported elsewhere [15]. THz-TDS measurements across 0.2 to 1.2 THz bandwidth at temperatures down to 2 K were performed in the transmission geometry using a custom-built spectrometer (Teravil Ltd./Ekspla uab, Lithuania) coupled to a Cryo-free Spectromag CFSM7T-1.5 magneto-optical cryostat (Oxford Instruments, UK). Broadband THz transients were generated and coherently detected using photoconductive antennas driven by a 1030 nm diode-pumped solid-state femtosecond laser (96 fs pulse duration, 76 MHz repetition rate). The entire optical path was continuously purged with dry nitrogen to eliminate absorption due to atmospheric moisture. A motorized low temperature insert enabled precise positioning of the thin-film sample, a reference substrate and a direct-bem aparture in the THz beam path. The time-domain signals were appropriately delayed and truncated to eliminate multiple reflections arising from the substrate and cryostat windows. Fast Fourier transforms were applied to extract the frequency-dependent amplitude and phase. To isolate the film's optical properties, the complex transmission through the reference was first determined experimentally relative to the direct beam to extract its frequency dependent refractive index ($n$) and extinction coefficient ($k$). Subsequently, the complex transmission through the sample was obtained as a ratio with respect to the reference substrate. Theoretical transfer functions for both the sample and reference were obtained using the Fresnel equations [19]. The intrinsic complex optical constants of the $Ti_{40}V_{60}$ thin films were extracted by minimizing the error between the experimental and theoretical transfer functions at each frequency, utilizing a Nelder-Mead simplex optimization algorithm implemented in Python [17].

As the 20 nm film thickness is deeply sub-wavelength, the transmitted field modifications are directly governed by the complex sheet conductivity [20]. Figs. 1(a)-(b) and 2(a)-(b) present the three-dimensional (3D) plots of the real ($\sigma_1$) and imaginary ($\sigma_2$) components of optical conductivity ($\sigma = \sigma_1 + i\sigma_2$) for TiV-1 and TiV-2 respectively in a temperature range spanning from the superconducting state (2 K) up to the normal state (10 K). The 3D plots of $\sigma_2$ in Fig. 1(b) and 2(b) are presented with reversed temperature and frequency axes to improve the clarity. At temperatures below the $T_C$, the real part of the complex optical conductivity is suppressed at low frequencies. This indicates a reduction in dissipative quasiparticle transport and the opening of the superconducting energy gap as the normal electrons condense into the coherent ground state [21]. The residual quasiparticle absorption near the gap edge remains pronounced due to gap smearing effects typical of the dirty-limit systems [22]. At the same time, the imaginary part of the conductivity captures the inductive response of the emerging Cooper pairs. Above $T_C$, the $\sigma_2$ remains extremely small. Notably, in the high-frequency normal-state regime of TiV-1 and TiV-2, the $\sigma_2$ exhibits a negative contribution. Such behaviour has also been reported in the highly disordered NbN films and is attributed to disorder-induced localization effects [23]. The finite-frequency THz field probes the enhanced dielectric polarizability of these developing localized states, confirming that normal-state transport is governed by defect scattering [23]. However, below the $T_C$, this subtle normal-state response is entirely overwhelmed by the superconducting quantum coherence. $\sigma_2$ rises sharply, developing a steep $1/_{\omega}$ frequency dependence. This inductive buildup is the direct electrodynamic signature of the resistance-less acceleration of the Cooper pairs [23], forming the physical basis for the high $L_K$ required by the resonator-based quantum sensors. Figs. 1(c) and 2(c) represent the low-energy complex conductivity at 0.2 THz, explicitly capturing the redistribution of the electrodynamic response across the superconducting transition. The sharp suppression of dissipative quasiparticle transport near the transition is perfectly mirrored by a steep, saturating rise in the inductive Cooper pair response.

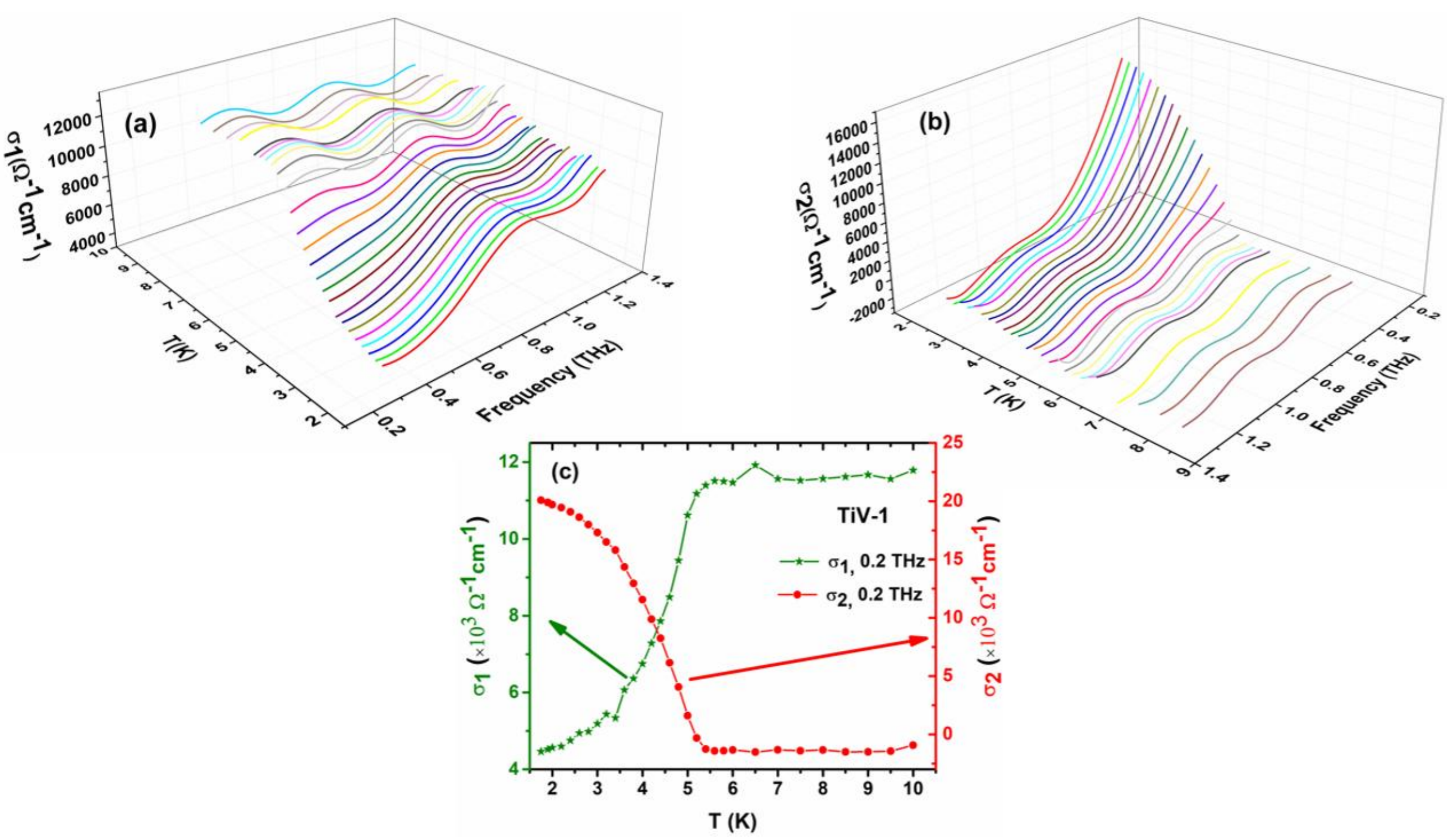

**Fig. 1 (a)** *Three-dimensional plots of the real part of the THz conductivity for the TiV-1 thin film. The dip in the low-frequency region below the* $T_C$ *indicates a superconducting gap in the quasiparticle states.* ***(b)*** *The imaginary part of the conductivity as a function of frequency exhibits* *a 1/ω dependence and becomes negative at higher frequencies.* ***(c)*** *Temperature dependence of the real and imaginary parts of THz conductivity of TiV-1 measured at 0.2 THz.*

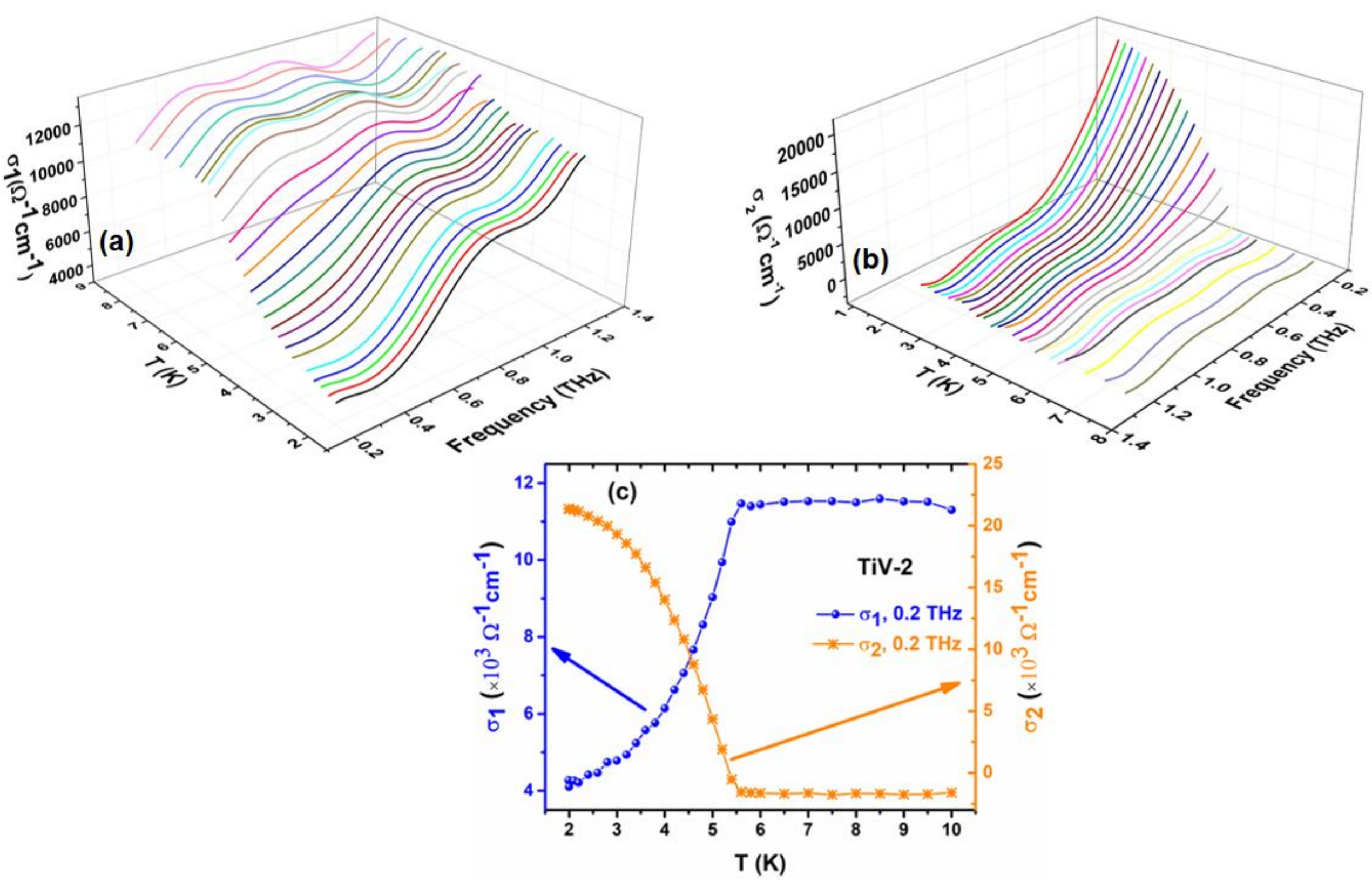

**Fig. 2 (a)** *Three-dimensional plots of the real part of the THz conductivity for the TiV-2 thin film. The dip in the low-frequency region below the $T_C$ indicates a superconducting gap in the quasiparticle states.* ***(b)*** *The Imaginary part of the conductivity as a function of frequency exhibits a 1/ω dependence and becomes negative at higher frequencies.* ***(c)*** *Temperature dependence of the real and imaginary parts of THz conductivity of TiV-2 measured at 0.2 THz.*

In the disordered thin films, electronic transport is dominated by the scattering effects [24,25]. The electrodynamic response of conventional superconductors in the dirty limit can be effectively described using the Mattis-Bardeen (M-B) formalism, which is based on the framework of BCS theory [26]. In the superconducting state, the frequency-dependent complex optical conductivity of the present films is expected to follow this formalism. To normalize the influence of the single electron state matrix elements, and to account for the localization effects, the conductivity in the superconducting state is commonly normalized by the normal-state conductivity measured slightly above the $T_C$. Within this framework, the real and imaginary parts of the optical conductivity are described by the standard M-B integral expressions (1) and (2), which incorporate the superconducting energy gap Δ, the photon energy $\hbar\omega$, and the Fermi-Dirac distribution function *f(E,T)*. For photon energies exceeding 2Δ, the lower integration limit is modified accordingly.

$$\frac{\sigma_1(\omega,T)}{\sigma_n} = \frac{2}{\hbar\omega}\int_{\Delta}^{\infty} \frac{[f(\varepsilon)-f(\varepsilon+\hbar\omega)](\varepsilon^2+\Delta^2+\hbar\omega\varepsilon)}{(\varepsilon^2-\Delta^2)^{\frac{1}{2}}[(\varepsilon+\hbar\omega)^2-\Delta^2]^{\frac{1}{2}}} d\varepsilon + \frac{1}{\hbar\omega}\int_{\Delta-\hbar\omega}^{-\Delta} \frac{[1-2f(\varepsilon+\hbar\omega)](\varepsilon^2+\Delta^2+\hbar\omega\varepsilon)}{(\varepsilon^2-\Delta^2)^{\frac{1}{2}}[(\varepsilon+\hbar\omega)^2-\Delta^2]^{\frac{1}{2}}} d\varepsilon \quad (1)$$

$$\frac{\sigma_2(\omega,T)}{\sigma_n} = \frac{1}{\hbar\omega}\int_{\Delta-\hbar\omega,-\Delta}^{\Delta} \frac{[1-2f(\varepsilon+\hbar\omega)](\varepsilon^2+\Delta^2+\hbar\omega\varepsilon)}{(\varepsilon^2-\Delta^2)^{\frac{1}{2}}[(\varepsilon+\hbar\omega)^2-\Delta^2]^{\frac{1}{2}}} d\varepsilon \quad (2)$$

As shown in the Figs. 3(a) and 3(d), the normalized real part of conductivity $\sigma_1/\sigma_n$, exhibits a systematic low-frequency suppression below the $T_C$ for both TiV-1 and TiV-2, consistent with superconducting gap formation and reduced quasiparticle dissipation. The corresponding M-B fits at 2 K are presented in the Figs. 3(b)-(c) and 3(e)-(f). The M-B formalism reproduces the overall spectral evolution reasonably well, particularly the suppression of $\sigma_1$ and the inductive enhancement of $\sigma_2$ at intermediate and high frequencies, yielding energy gaps of 1.86 and 1.9 meV for TiV-1 and TiV-2 respectively. However, noticeable deviations from M-B formalism are observed in the low-frequency sub-gap regime. In the Figs. 3(b) and 3(e), a finite residual $\sigma_1$ persists even at 2 K, whereas an ideal fully gapped superconductor is expected to exhibit nearly vanishing conductivity below the gap edge. Such residual sub-gap absorption is commonly observed in the strongly disordered superconductors and is generally attributed to impurity-induced quasiparticle states, spatial inhomogeneity, and disorder-driven broadening of the quasiparticle density of states [27,28]. Similarly, the experimentally measured $\sigma_2/\sigma_n$ in Figs. 3(c) and 3(f) shows weaker low-frequency enhancement than predicted by the ideal M-B model, probably originating from strong scattering and localization effects not fully captured within the conventional dirty-limit description [29]. Similar deviations from the ideal M-B formalism have been reported in disordered NbN and NbTiN thin films [23,28].

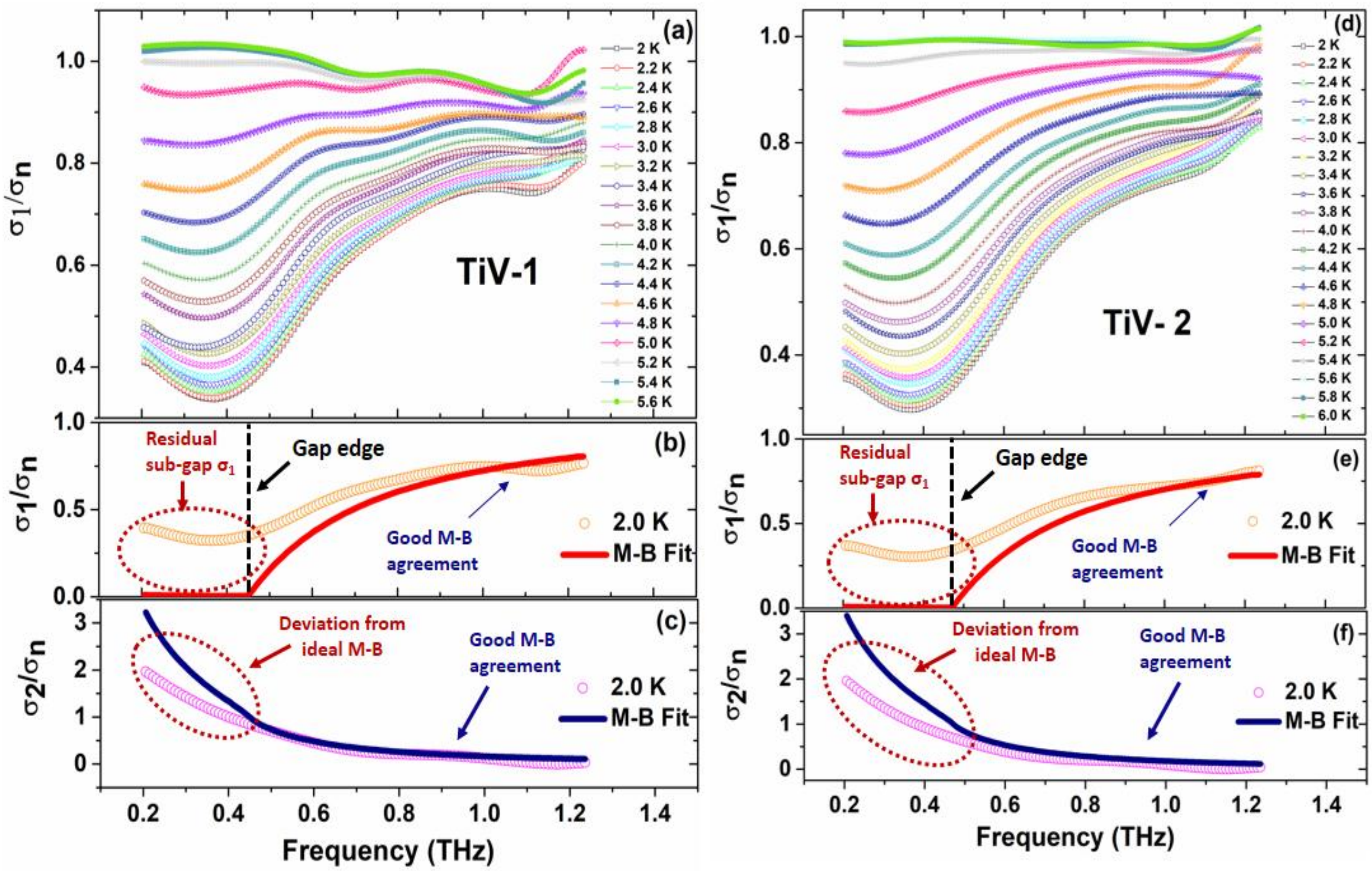

**Fig. 3** ***(a) and (d)*** *Frequency dependence of the normalized real part of complex conductivity at different temperatures below the $T_C$ for TiV-1 and TiV-2 respectively.* ***(b)*** *and* ***(e)*** *Normalized real part of complex conductivity of TiV-1 and TiV-2 respectively, at 2 K, along with the M-B fits.* ***(c)*** *and* ***(f)*** *Normalized imaginary part of complex conductivity of TiV-1 and TiV-2 respectively at 2 K, along with the corresponding M-B fit.*

The microscopic energy scales of the condensate were evaluated by analyzing the frequency-dependent complex conductivity. To precisely determine the zero-temperature energy gap $\Delta(0)$ and evaluate the macroscopic phase stiffness, we analyzed the temperature evolution of the penetration depth extracted directly from the imaginary part of conductivity at 0.2 THz using equation (3) (see Fig. 4).

$$\lambda = \frac{c}{\sqrt{4\pi\omega\sigma_2}} \tag{3}$$

Here, $c$ is the speed of light in vacuum and $\omega$ is the frequency of the radiation [21]. As the films are disordered, the dirty-limit BCS expression [equation (4)]_provides an excellent description of the experimental $\lambda^{-2}(T)$ data [Figs. 4(a) and (b)] for both the samples, where $\Delta(0)$ is the fitting parameter [30].

$$\frac{\lambda^{-2}(T)}{\lambda^{-2}(0)} = \frac{\Delta(T)}{\Delta(0)} tanh\left[\frac{\Delta(T)}{2k_B T}\right] \qquad (4)$$

By fitting equation (4), the $\Delta(0)$is found to be 0.91 ± 0.007 meV for TiV-1 and 1.03 ± 0.007 meV for TiV-2, alongside the $\lambda(0)$ values of 550.5 ± 0.7 nm and 532.1 ± 0.5 nm, respectively. These gap values are in good agreement with those extracted independently from the M-B analysis of the complex conductivity presented with Fig. 3. Using these precisely fitted gap values and the measured $T_C$ (5.72 K for TiV-1, 5.9 K for TiV-2), the resulting coupling ratios $\left(\frac{2\Delta(0)}{K_B T_C}\right)$ are found to be 3.69 and 4.05. The slightly lower value of $\lambda(0)$ and higher energy gap in TiV-2 as compared to TiV-1 indicate a stronger superconducting pairing strength in the TiV-2 film. These coupling ratios clearly exceed the standard BCS weak-coupling limit of 3.53. Similar to the findings in strongly disordered transition metal nitrides [4], this enhancement confirms that the disordered Ti-V lattice resides in the strong-coupling regime, which makes the superconducting state significantly more robust against thermal fluctuations.

The temperature dependence of the penetration depth was also analyzed using the model proposed by Gorter and Casimir (G-C). In this approach, the superconducting charge carriers are divided into normal and superfluid components, whose relative fractions vary with temperature. The penetration depth as a function of temperature was fitted using equation (5), which describes the reduction of superfluid density with increasing temperature and its complete disappearance at the $T_C$. The fitting of the $\lambda(T)$ data is shown in fig. 4(c) and (d) for TiV-1 and TiV-2 respectively.

$$\lambda(T) = \lambda(0)\left[1 - \left(\frac{T}{T_C}\right)^{\alpha}\right]^{-1/2} \qquad (5)$$

In the G-C two-fluid model, the exponent is conventionally taken as 4 [4]. However, the experimental data for the present films are better described using $\alpha$ = 4.13 for TiV-1 and $\alpha$ = 3.7 for TiV-2, indicating slight deviations from the ideal two-fluid behavior. Similar deviation from the conventional $\alpha = 4$ dependence have also been reported in other disordered superconducting films [4,31]. From the fits, the values of $\lambda(0)$ were estimated to be 561 nm and 524 nm for TiV-1 and TiV-2 respectively. These values are consistent with the values obtained from the dirty-limit BCS analysis presented earlier. The penetration depth increases slowly at low temperatures and rises rapidly as the temperature approaches the $T_C$, reflecting the decrease in superfluid density. The agreement between the experimental data and the two-fluid model indicates that the temperature dependence of superfluid density in these films follows the conventional behavior of disordered superconductors. These electrodynamic parameters ultimately define how well these films will work in quantum devices. From the penetration depth, the zero-temperature superfluid densities estimated using equation (6) [32] are found to be $9.28 \times 10^{25}\, m^{-3}$ and $9.98 \times 10^{25}\, m^{-3}$ for TiV-1 and TiV-2 respectively. The condensate density directly gives the sheet $L_K$ [equation (7)] [33] as 19.15 pH/sq and 17.81 pH/sq respectively.

$$n_s = \frac{m_e}{\mu_0 e^2 \lambda^2} \tag{6}$$

$$L_k = \frac{m_e}{n_s e^2} \frac{l}{A} \tag{7}$$

$$J = 0.62 \frac{d}{\lambda^2} \tag{8}$$

The observed enhancement in $L_K$ and slight suppression of $T_C$ with increasing deposition pressure can be directly linked to the thermodynamics of the sputtering process. At a higher deposition pressure (0.7 $\mu$bar for TiV-1 versus 0.63 $\mu$bar for TiV-2), the sputtered Ti and V atoms experience increased collision with the argon sputtering gas. This thermalization reduces the kinetic energy of the adatoms upon reaching the substrate, which inherently restricts the surface mobility during film growth [15]. Consequently, the higher-pressure deposition promotes a more granular microstructure with a higher density of lattice defects and grain boundaries. This pressure induced structural disorder increases the normal-state electron scattering rate, pushing the Ti-V matrix deeper into the dirty limit. This suppression of the electronic mean free path fundamentally drives the reduction in superfluid density [27] and the corresponding enhancement of the macroscopic kinetic inductance.

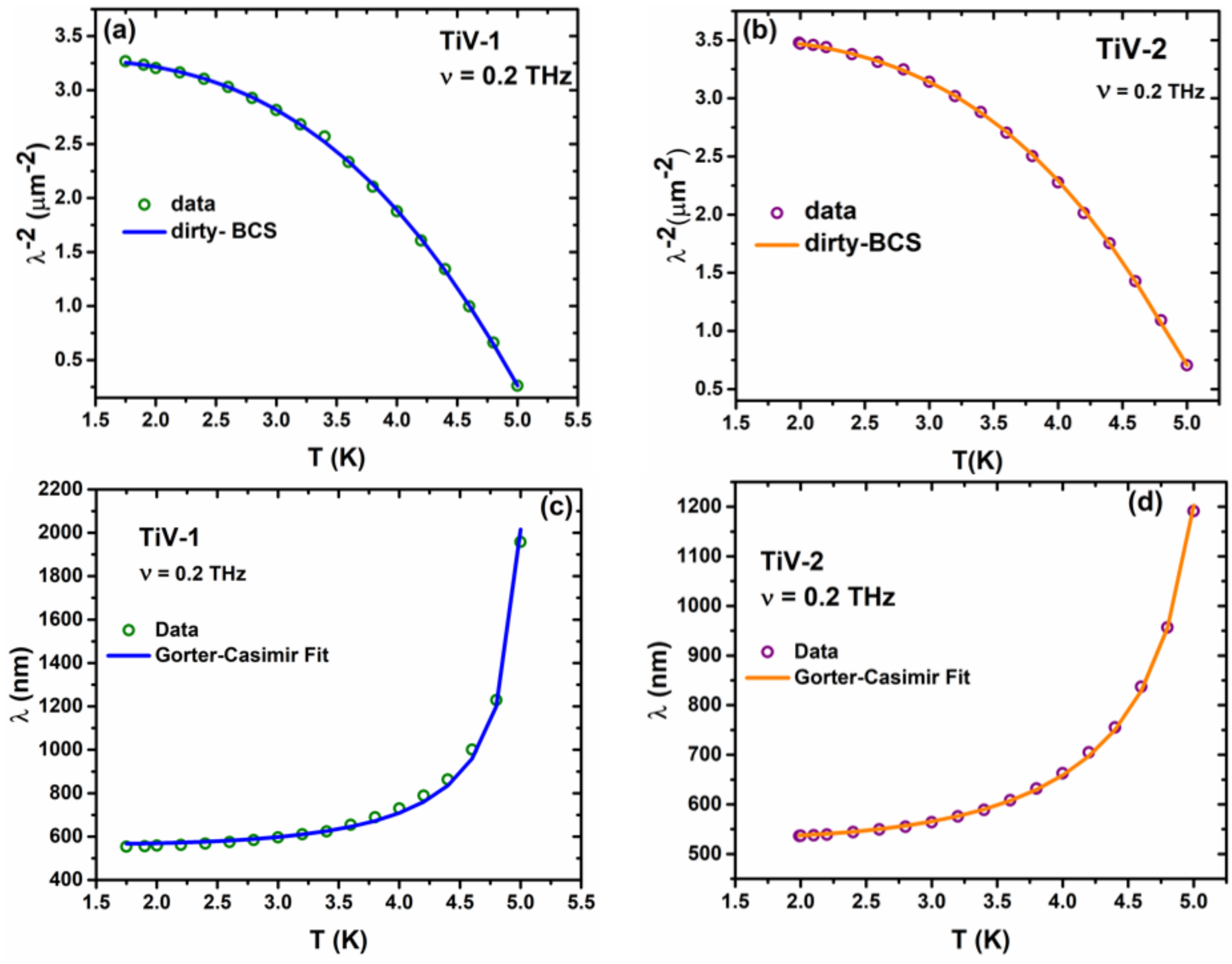

**Fig. 4** *Temperature dependence of the magnetic penetration depth:* ***(a, b)*** *The inverse square of the penetration depth, $\lambda^{-2}(T)$, evaluated at 0.2 THz for TiV-1 and TiV-2, respectively. The solid lines denote theoretical fits to the experimental data using the dirty-limit BCS formulation.* ***(c, d)*** *The absolute penetration depth plotted as a function of temperature for TiV-1 and TiV-2. The solid lines represent fits based on the empirical Gorter-Casimir two-fluid model.*

To confirm that this highly inductive state is stable, we estimated the superfluid stiffness using equation (8) as 35.1 and 37.8 meV for TiV-1 and TiV-2 respectively. All the estimated values [$\lambda(0)$, $\Delta(0)$, $L_K$, $n_s$, and $J$] for both the samples are presented in table 1. The finding that $J$(2 K) is much larger than the pairing gap $\Delta(0)$ proves that the phase of the superconducting order parameter is rigid. This means that the transition at $T_C$ occurs because of the loss of pairing amplitude and not the phase fluctuations. This stability is a critical requirement for low-noise applications like KIDs. The present results show that the localization and scattering effects in the Ti-V alloys can be controlled during deposition to achieve the large stable $L_K$ needed for the next-generation quantum sensors.

***Table 1*** *Characteristic material properties of superconducting $Ti_{40}V_{60}$ thin films (TiV-1 and TiV-2) obtained from complex optical conductivity*

| S. No. | Sample | $\lambda$ (0) (Dirty-BCS) nm | $\Delta$(0) meV | $n_s$ ($\times 10^{25} m^{-3}$) | $L_{kin}$ (pH/sq) | $J$ (meV) |
|---|---|---|---|---|---|---|
| **1.** | TiV-1 | 550.5 ± 0.7 | 0.91 ± 0.007 | 9.28 | 19.15 | 35.1 |
| **2.** | TiV-2 | 532.1 ± 0.5 | 1.03 ± 0.007 | 9.98 | 17.81 | 37.8 |

In conclusion, we have probed the THz electrodynamics and inductive response of the disordered superconducting $Ti_{40}V_{60}$ alloy thin films. Through temperature and frequency dependent THz-TDS, we established that these metallic alloys operate in the dirty limit, exhibiting strong-coupling superconductivity ($\frac{2\Delta}{K_B T_C} \sim 3.65 - 4.09$) and sub-gap quasiparticle absorption characteristics of structural disorder. Despite this defect scattering, the macroscopic quantum condensate remains remarkably robust. We demonstrate that the emergent state of Cooper pairs yields stable, high sheet $L_K$ in the range of 17 to 19 pH/sq. Furthermore, the extracted superfluid stiffness significantly exceeds the zero-temperature pairing gap, confirming a rigid superconducting phase where the thermal transition is governed entirely by Cooper pair breaking rather than phase fluctuations. Our results establish that adjusting the deposition pressure provides a direct, stoichiometry-free tuning knob to control the localization effects, structural disorder, and the resulting inductive response. By circumventing the complexities inherent to conventional nitride growth, the Ti-V alloys offer a highly controllable fabrication pathway. These findings highlight the compelling physical properties of disordered Ti-V thin films and position them as a robust, scalable, and highly competitive materials platform for next-generation, high-responsivity quantum sensors, including KIDs and SNSPDs.

# Data availability statement

The data that supports the findings of this study are available from the corresponding author upon reasonable request.

# References

[1] C. Zhang, W. Zhang, J. Huang, L. You, H. Li, C. lv, T. Sugihara, M. Watanabe, H. Zhou, Z. Wang, and X. Xie, AIP Advances **9,** 075214 (2019).

[2] F. Mazzocchi, D. Strauß, and T. Scherer, in *First measurements of NbN based KIDs for polarimetric plasma diagnostics*, 1 (2020).

[3] R. Steven de, R. Thomas, K. Kevin, A. Yannic, C. Tonny, B. Jochem, and V. Pieter de, in *Single photon response of MKIDs made of disordered superconductors*, PC1310304 (2024).

[4] O. Saritas, F. Bolle, Y. Lin, M. Dressel, R. Potjan, M. Wislicenus, A. Reck, and M. Scheffler, Applied Physics Letters **127,** 192601 (2025).

[5] A. V. Semenov, I. A. Devyatov, M. P. Westig, and T. M. Klapwijk, Physical Review Applied **13,** 024079 (2020).

[6] F. Avitabile, F. Colangelo, M. Y. Mikhailov, Z. Makhdoumi Kakhaki, A. Kumar, I. Esmaeil Zadeh, C. Attanasio, and C. Cirillo, Applied Physics Letters **127,** 172601 (2025).

[7] K. Waki, T. Yamashita, S. i. Inoue, S. Miki, H. Terai, R. Ikuta, T. Yamamoto, and N. Imoto, in *Characterization of NbTiN-Based Superconducting Nanowire Single-Photon Detectors on Si Waveguide*, 1 (2015).

[8] A. Engel, A. Aeschbacher, K. Inderbitzin, A. Schilling, K. Il'in, M. Hofherr, M. Siegel, A. Semenov, and H. W. Hübers, Applied Physics Letters **100,** 062601 (2012).

[9] M. Erbe, R. Berrazouane, S. Geyer, L. Stasi, F. van der Brugge, G. Gras, M. Schmidt, A. D. Wieck, A. Ludwig, F. Bussières, and R. J. Warburton, Physical Review Applied **22,** 014072 (2024).

[10] I. Charaev, Y. Morimoto, A. Dane, A. Agarwal, M. Colangelo, and K. K. Berggren, Applied Physics Letters **116,** 242603 (2020).

[11] J. S. Luskin, E. Schmidt, B. Korzh, A. D. Beyer, B. Bumble, J. P. Allmaras, A. B. Walter, E. E. Wollman, L. Narváez, V. B. Verma, S. W. Nam, I. Charaev, M. Colangelo, K. K. Berggren, C. Peña, M. Spiropulu, M. Garcia-Sciveres, S. Derenzo, and M. D. Shaw, Applied Physics Letters **122,** 243506 (2023).

[12] H. Le Jeannic, V. B. Verma, A. Cavaillès, F. Marsili, M. D. Shaw, K. Huang, O. Morin, S. W. Nam, and J. Laurat, Optics Letters **41,** 5341 (2016).

[13] S. C. Pandey, S. Sharma, K. K. Pandey, P. Gupta, S. Rai, R. Singh, and M. K. Chattopadhyay, Journal of Applied Physics **137,** 113902 (2025).

[14] S. C. Pandey, S. Sharma, P. Gupta, L. S. Sharath Chandra, and M. K. Chattopadhyay, Journal of Physics and Chemistry of Solids **216,** 113761 (2026).

[15] S. C. Pandey, S. Sharma, R. Venkatesh, A. Khandelwal, L. S. Sharath Chandra, and M. K. Chattopadhyay, Superconductor Science and Technology **39,** 025008 (2026).

[16] S. C. Pandey, S. Sharma, A. Khandelwal, and M. K. Chattopadhyay, AIP Conference Proceedings **3198,** 020113 (2025).
[17] S. Sharma, A. Khandelwal, E. P. Amaladass, A. S, R. Sk, J. J, A. Mani, and M. K. Chattopadhyay, Journal of Applied Physics **128,** 183901 (2020).
[18] J. Neu and C. A. Schmuttenmaer, Journal of Applied Physics **124,** 231101 (2018).
[19] N. Kida, M. Hangyo, and M. Tonouchi, Physical Review B **62,** R11965 (2000).
[20] M. Tinkham, *Introduction to superconductivity*, 2nd ed. ed. (Dover Publications, Mineola, N.Y, 2004).
[21] M. Dressel, Advances in Condensed Matter Physics **2013,** 104379 (2013).
[22] T. Kubo, Physical Review Applied **17,** 014018 (2022).
[23] B. Cheng, L. Wu, N. J. Laurita, H. Singh, M. Chand, P. Raychaudhuri, and N. P. Armitage, Physical Review B **93,** 180511 (2016).
[24] S. K. Maiti, International Journal of Nanoscience **07,** 171 (2008).
[25] V. Bagwe, R. Duhan, B. Chalke, J. Parmar, S. Basistha, and P. Raychaudhuri, Physical Review B **109,** 104519 (2024).
[26] D. C. Mattis and J. Bardeen, Physical Review **111,** 412 (1958).
[27] P. Raychaudhuri and S. Dutta, Journal of Physics: Condensed Matter **34,** 083001 (2022).
[28] T. Hong, K. Choi, K. Ik Sim, T. Ha, B. Cheol Park, H. Yamamori, and J. Hoon Kim, Journal of Applied Physics **114,** 243905 (2013).
[29] G. Seibold, L. Benfatto, and C. Castellani, Physical Review B **96,** 144507 (2017).
[30] A. Kamlapure, M. Mondal, M. Chand, A. Mishra, J. Jesudasan, V. Bagwe, L. Benfatto, V. Tripathi, and P. Raychaudhuri, Applied Physics Letters **96,** 072509 (2010).
[31] K. Hashimoto, Y. Mizukami, R. Katsumata, H. Shishido, M. Yamashita, H. Ikeda, Y. Matsuda, J. A. Schlueter, J. D. Fletcher, A. Carrington, D. Gnida, D. Kaczorowski, and T. Shibauchi, Proceedings of the National Academy of Sciences **110,** 3293 (2013).
[32] M. Dressel, G. Gruener, and G. F. Bertsch, American Journal of Physics **70,** 1269 (2002).
[33] R. Meservey and P. M. Tedrow, Journal of Applied Physics **40,** 2028 (1969).